\documentclass[preprint,12pt]{elsarticle}

\usepackage{graphicx}
\usepackage{dcolumn}
\usepackage{bm}
\usepackage{amssymb}
\usepackage{amsmath}
\usepackage{color}

\journal{https://www.105groupscience.com/}

\begin{document}

\begin{frontmatter}

\title{Unpredictability in seasonal infectious diseases spread}

\author{Enrique C. Gabrick$^{1,*}$, Elaheh Sayari$^{1}$, Paulo R. Protachevicz$^{2,3}$, Jos\'e D. Szezech
Jr.$^{1,4}$, Kelly C. Iarosz$^{5,6}$, Silvio L.T. de Souza$^7$, Alexandre C.L.
Almeida$^8$, Ricardo L. Viana$^{2,9}$, Iber\^e L. Caldas$^2$, Antonio M.
Batista$^{1,4}$}
\address{$^1$Graduate Program in Science, State University of Ponta Grossa,
84030-900, Ponta Grossa, PR, Brazil.\\}
\address{$^2$Institute of Physics, University of S\~ao Paulo, S\~ao Paulo, SP,
Brazil.\\}
\address{$^3$Institute for Complex Systems and Mathematical Biology, University of 
Aberdeen, Aberdeen, United Kingdom.\\}
\address{$^4$Department of Mathematics and Statistics, State University of Ponta
Grossa, 84030-900, Ponta Grossa, PR, Brazil.\\}
\address{$^5$University Center UNIFATEB,  84266-010, Tel\^emaco Borba, PR,
Brazil.\\}
\address{$^6$Graduate Program in Chemical Engineering Federal Technological
University of Paran\'a, Ponta Grossa, PR, Brazil.\\}
\address{$^7$Federal University of S\~ao Jo\~ao del-Rei, Campus Centro-Oeste,
35501-296, Divin\'opolis, MG, Brazil.\\}
\address{$^8$Physics and Mathematics Department, Federal University of S\~ao
Jo\~ao del-Rei, 36415-000, Ouro Branco, MG, Brazil.\\}
\address{$^9$Department of Physics, Federal University of Paran\'a, Curitiba,
PR, Brazil.\\}

\cortext[cor]{ecgabrick@gmail.com}

\date{\today}

\begin{abstract}
In this work, we study the unpredictability of seasonal infectious
diseases considering a SEIRS model with seasonal forcing. To investigate 
the dynamical behaviour, we compute bifurcation diagrams type hysteresis and
their respective Lyapunov exponents. Our results from bifurcations and the 
largest Lyapunov exponent show bistable dynamics for all the parameters of 
the model. Choosing the inverse of latent period as control parameter,
over 70\% of the interval comprises the coexistence of
periodic and chaotic attractors, bistable dynamics. Despite
the competition between these attractors, the chaotic ones are preferred.
The  bistability occurs in two wide regions.
One of these regions is limited by periodic attractors, while periodic 
and chaotic attractors bound the other.  As the boundary of the second
bistable region is composed of periodic and chaotic attractors, it is possible
to interpret these critical points as tipping points. In other words, depending on the
latent period, a periodic attractor (predictability) can evolve to a chaotic
attractor (unpredictability). Therefore, we show that unpredictability 
is associated with bistable dynamics preferably chaotic, and,  furthermore, 
there is a tipping point associated with unpredictable dynamics.
\end{abstract}


\end{frontmatter}


\section{Introduction}

The study of the spread of diseases is an important interdisciplinary 
research topic \cite{may1991}. Mathematical models are essential to understanding, forecasting,
and studying control measures for infectious diseases spread \cite{physicaa}. In
general, the epidemic models are compartmental, i.e., they divide the host
population ($N$) into compartments, for instance susceptible ($S$), exposed
($E$), infected ($I$), and recovered ($R$) \cite{rbef}. $S$ is related to 
 healthy individuals who can contract the disease. $E$ corresponds to the
individuals in latent \cite{sharma21} and/or incubation period \cite{quanxing}.
In the latent period, the individuals can not transmit the disease
\cite{physicaa}. In the incubation period, the exposed can transmit the disease
with a lower incidence than the infected individuals \cite{amaku21b, seir}. $I$ is
associated with individuals who transmit the disease. $R$ is related to the 
individuals who were infected and got immunity, permanent \cite{physicaa} or
temporary \cite{michele}. The composition of these compartments forms the
classical epidemics models: Susceptible-Infected (SI) \cite{abdelaziz2018},
Susceptible-Infected-Susceptible (SIS) \cite{nakamura2019},
Susceptible-Infected-Recovered (SIR) \cite{wei2021},
Susceptible-Infected-Recovered-Susceptible (SIRS) \cite{wang2021},
Susceptible-Exposed-Infected-Recovered (SEIR) \cite{silvio, physicaa}, and
Susceptible-Exposed-Infected-Recovered-Susceptible (SEIRS) 
\cite{michele}. An
introduction to these models can be found in Ref. \cite{rbef}. These models have
been used to study the dynamic of many diseases, for example, COVID-19
\cite{cooper2020}, dengue fever \cite{aguiar2008}, and childhood epidemics
(e.g., measles, diphtheria, and chickenpox) \cite{olsen1990}. 

Some of these
diseases have seasonal behaviour, like measles, chickenpox, pertussis, and
others \cite{tanaka2013, galvis2022}. The common characteristic of seasonal
diseases is the recurrence of new outbreaks after a period of time. The 
motivation to work with seasonal models is to predict future outbreaks and
propose control measures \cite{moneim2005}. 

Seasonal models have been introduced since 1928 \cite{buonomo2017}. They present a
rich variety of oscillatory phenomena \cite{keeling2001}. London and Yorke
\cite{london1973} considered an epidemic model with seasonally 
varying contact rates forcing, studied the recurrent
outbreaks of measles, chickenpox, and mumps in New York City. 
Their simulations reproduced the observed pattern in annual outbreaks of 
chickenpox, mumps and biennial outbreaks of measles and also verified that
the mean contact rate is higher in winter than in summer.
Considering a SEIR model with seasonal components, Olsen and Schaffer
\cite{olsen1990} showed, from real data, that measles epidemics are
inherently chaotic.  Aguiar et al. \cite{aguiar2011} analysed
a seasonally forced SIR epidemic model for dengue fever with temporary 
cross-immunity and the possibility of secondary infection. Their results showed that
the addition of seasonal forcing induces chaotic dynamics, which is related to
the decrease in predictability. Similar results in a SIR model were reported
by Stollenwerk et al. \cite{stollenwerk2022}. He et al. \cite{he2020} explored
a SEIR epidemic model based on the COVID-19 data from Hubei province. With the
introduction of seasonality and stochastic infection, the model becomes nonlinear with chaos. 
In addition to this work,  from the analysis of epidemiological 
data from 14 countries, the work of Jones and Strigul \cite{jones2020} 
suggested that the COVID-19 spread is chaotic. Bilal et al. \cite{bilal2016} 
studied changes in the bifurcations of the seasonally
forced SIR model considering a transmission rate modulated temporally. 
By analysing the bifurcation diagrams and respective Lyapunov exponents, in the forward 
and backward directions of the strength of seasonality, their results
showed the coexistence between periodic and chaotic attractors, known as
bistability dynamics. Bistability in an epidemic model also was reported 
by Ventura et al. \cite{ventura2022}. They considered a model with mobility where the spreading of disease 
occurs in temporal networks of mobile agents. In their model, they considered the movement
of susceptible in the oppositive direction of infected agents. By developing a
semi-analytic approach, they showed that the bistability is caused by the 
spatial emergence of susceptible clustering. 

Many natural processes exhibit multistability, i.e., the  asymptotic state
evolves to a large number of coexistence attractors for a fixed parameter set
\cite{cheng2022}. In these systems, the transient for the final attractor depends
strongly on the initial condition \cite{feudel1997,feudel2008}. The existence of
two alternative states is very important to climate science \cite{lenton2011},
ecology \cite{scheffer2015}, and epidemiology
\cite{wei2021,bilal2016,ventura2022}. When the multistability region is bounded
by contrasting attractors and an abrupt shift between these attractors occurs,
the threshold points in which this transition occurs are called 
tipping points \cite{dakos2019}. 

Tipping points are found in the process of desertification \cite{dakos2019},
cancer epidemiology \cite{wright2022}, Duffing oscillator \cite{everton},
epidemic models \cite{ansari2021, regan2020, francomano2018}, ecological models 
\cite{meng2020}, and others \cite{meng2021}. Mathematically, the tipping points
correspond to bifurcations \cite{dakos2019}, that, in general, have long
transient lifetimes \cite{grebogi1985}. In the ecological sense, long transients
were studied by Hastings et al. \cite{hastings2021}.

In seasonal disease spread, an unclear problem is the limit to forecast
precision for the outbreak, as observed by Scarpino and Petri \cite{scarpino2019}.
They studied the time series from ten different diseases (for example, dengue,
influenza, measles, and mumps) and demonstrated that the predictability
decreases when the time series length is increased. Furthermore, their results
showed that the forecast horizon varies by different illnesses. From the other
works, it is known that unpredictability is associated with the 
chaotic dynamics
\cite{stollenwerk2012}. However, only chaotic dynamics do not provide a
satisfactory answer to understanding the mechanism behind unpredictability, since
the chaotic attractors are predictable until a Lyapunov time in the order to the
inverse of the largest Lyapunov exponent.

Our main goal in this research is to study the mechanism 
behind the unpredictability in seasonal infectious diseases. In order to that, 
we consider a SEIR model with temporary immunity and seasonal forcing \cite{bai2012,nyi}.
Firstly, we show that the basic reproductive rate $R_0$ depends on the seasonality
parameters, such as seasonality degree and frequency. In sequence, we
consider numerical simulations which exhibit the existence of bistability for
all parameters in the model, which is characterised by the coexistence
of periodic and chaotic attractors. We verify this dynamical behaviour by
bifurcations diagram type hysteresis and the largest Lyapunov exponent.
Despite the rich dynamics  in all the parameters, we select the inverse of 
the latent period to study the unpredictability phenomenon. A wide
range of this parameter comprises diseases with short (hours) and large
(days) latent period. Our results show that the dynamics are sustained
over 70\% by bistability between chaotic and periodic attractors. 
This bistability appears in two large ranges. In the first one, the
probability of one initial condition evolving to the chaotic attractor is 51\%,
while in the second range is 63\%. Furthermore, these two ranges are delimited, in their
crisis points, by periodic and chaotic attractors (without bistability).
In this sense, it is possible to interpret these bifurcations as tipping points.
In this way, as novelty, we exhibit that the unpredictability in infectious disease
spread is associated with bistable dynamics and 
exists one tipping point associated with it.

Our work is organised as follows: In section 2, we present the model. Section 3
is dedicated for the study of bifurcations and crisis points. In Section 4, we
interpret the crisis points as tipping points. Finally, in Section 5, we draw
our conclusions.


\section{Model}

\begin{figure}[hbt]
\centering
\includegraphics[width=0.70\textwidth]{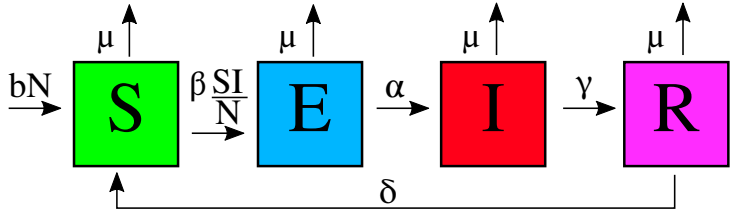}
\caption{Schematic representation of SEIRS model, where $N$ is the host
population, $b$ is the natural birth rate, $\mu$ is the natural death rate,
$\beta$ is the effective per capita contact rate, $\alpha$ is the rate at which
$E$ evolves to $I$, $\gamma$ is the recovered rate, $\delta$ is the rate in
which $R$ return to $S$ compartmental.}
\label{fig1}
\end{figure}

The SEIRS model divides the host population ($N$) into four compartments
\cite{physicaa,rbef,michele}: $S$, $E$, $I$, and $R$. A schematic representation
of the SEIRS model is shown in Fig. \ref{fig1}. The $R$ individuals lose  
immunity after a period time given by $1/\delta$ and return to $S$
compartmental. The model is given by
\begin{equation} 
\begin{split}
\frac{dS}{dt} & = bN(t)-\mu S(t)-\beta\frac{S(t)I(t)}{N(t)}+\delta R(t), \\
\frac{dE}{dt} & = \beta \frac{S(t)I(t)}{N(t)}-(\alpha+\mu)E(t), \\
\frac{dI}{dt} & = \alpha E(t)-(\gamma+\mu)I(t), \\
\frac{dR}{dt} & = \gamma I(t)-\mu R(t)-\delta R(t), \\
\label{eq1}
\end{split}
\end{equation}
where $b$ is the natural birth rate, $\mu$ is the natural death rate, $\alpha$
is the rate at which exposed individuals evolve to be infected, $\gamma$ is the 
recovered rate, $\delta$ is the rate at which the recovered individuals return
to the susceptible class after losing  immunity. The mean latent period is
given by  $1/\alpha$, the mean infectious period by $1/\gamma$, and the mean
time immunity by $1/\delta$. The force of infection is $\frac{\beta I}{N}$, 
where $\beta$ is the effective per capita contact rate of infective 
individuals and the incidence rate is $\frac{\beta S I}{N}$. 

We consider the transmission rate with a seasonal forcing given by
\begin{equation} 
\beta(t)=\beta_0 (1+\beta_1{\rm cos}\omega t),
\label{eq2}
\end{equation}
where $\beta_0$ is the average contact rate, $\beta_1$ ($0 \leq \beta_1 \leq 1$)
measures the seasonality degree, and $\omega$ is the frequency
\cite{olsen1990,bai2012,nyi}. Considering $\mu=b$, we obtain, from Eq. \ref{eq1}, $S+E+I+R=N$. Therefore, it
is possible, without loss of generality, to rewrite Eq. \ref{eq1} using 
the following
transformations: $s=S/N$, $e=E/N$, $i=I/N$, $r=R/N$. 

The equilibrium solutions of SEIRS
model are found by solving the following equations
\begin{equation} 
\begin{split}
0 & \equiv b-\mu s-\beta si+\delta r, \\
0 & \equiv \beta si-(\alpha+\mu)e, \\
0 & \equiv \alpha e-(\gamma+\mu)i, \\
0 & \equiv \gamma i-\mu r-\delta r,
\label{equilibrium}
\end{split}
\end{equation}
where the disease-free equilibrium is $(s^*,e^*,i^*,r^*)=(1,0,0,0)$, since $b=\mu$.
However, a very important equilibrium  solution is the endemic solution, which 
is given by
\begin{equation} 
\begin{split}
s^* & = \frac{(\alpha + \mu)(\gamma + \mu)}{\beta(t) \alpha} \equiv \frac{1}{R_0}, \\
e^* & = \frac{\gamma+\mu}{\alpha} \frac{\mu + \delta}{\beta(t)(\mu + \delta) - R_0 \delta \gamma}(bR_0 - \mu), \\
i^* & = \frac{\mu + \delta}{\beta(t)(\mu + \delta) - R_0 \delta \gamma}(bR_0-\mu), \\
r^* & = 1 - s^* - e^* - i^*,
\label{endemic}
\end{split}
\end{equation}
where $R_0$ is the basic reproductive ratio \cite{matt}. We consider $\mu=b$ to obtain a fixed population size.
If $\delta = 0$ and $\mu=b$, we recover the expression $i^* = \frac{\mu}{\beta}(R_0 - 1)$
that is the endemic fixed point for the SEIR model, as shown in Ref. \cite{matt}
without seasonal forcing.
In this way, the terms $\delta$, $b$, $\gamma$, $\beta_0$, and $\beta_1$
appear as correction terms in the infected individuals. 
$R_0 = R_0 (t) \propto \beta_0 (1+\beta_1{\rm cos}\omega t)$
is proportional to the seasonal parameters.

\begin{figure}[hbt]
\centering
\includegraphics[scale=0.45]{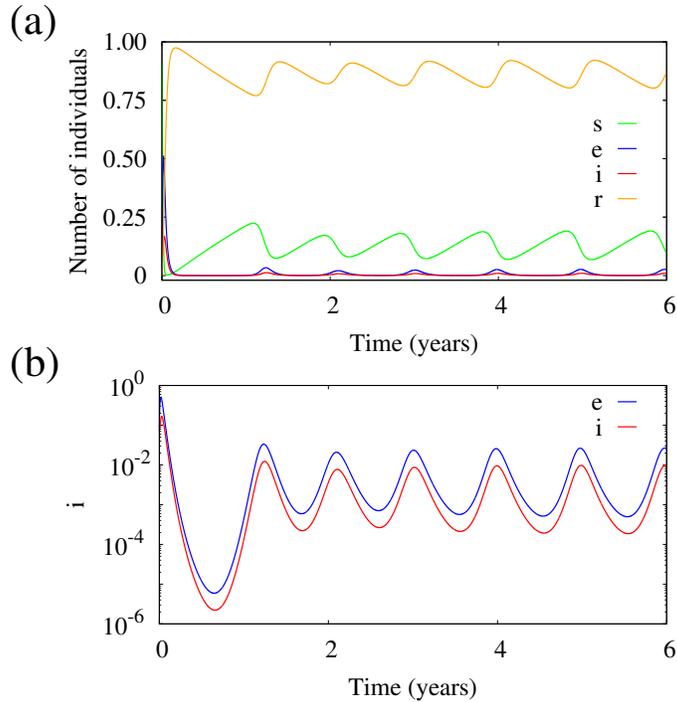}
\caption{(a) SEIRS time series. (b) Magnification in log scale of the exposed
(blue) and infected (red). We consider $\mu=0.02$, $\omega=2\pi$,
$\alpha=37.35$, $\gamma=100$, $\delta=0.25$, $\beta_1=0.15$, and $\beta_0=800$.}
\label{fig2}
\end{figure}

 However, Eq. \ref{eq2}
makes the system become nonautonomous. To build an autonomous system, we
introduce a new variable $T=\omega t$ and a new differential equation, that is
given by $\frac{dT}{dt}=\omega$. With these considerations, the equations become
\begin{equation} 
\begin{split}
\frac{ds}{dt} & = \mu-\mu s-\beta(T)si+\delta r, \\
\frac{de}{dt} & = \beta(T)si-(\alpha+\mu)e, \\
\frac{di}{dt} & = \alpha e-(\gamma+\mu)i, \\
\frac{dT}{dt} & = \omega,\\
r & = 1-s-e-i,
\label{eq3}
\end{split}
\end{equation}
where the differential equation for $r$ is replaced by the constraints. As an
example, a solution of Eq. \ref{eq3} is shown in Fig. \ref{fig2}(a) for all
variables. Differently from the standard SEIRS, the seasonal forcing produces
oscillations in the epidemic curves in accordance with $\omega$. The exposed and
infected curves are amplified in Fig. \ref{fig2}(b), in log scale. This result
shows a solution like a forced damped oscillator.

\begin{figure}[hbt]
\centering
\includegraphics[scale=0.23]{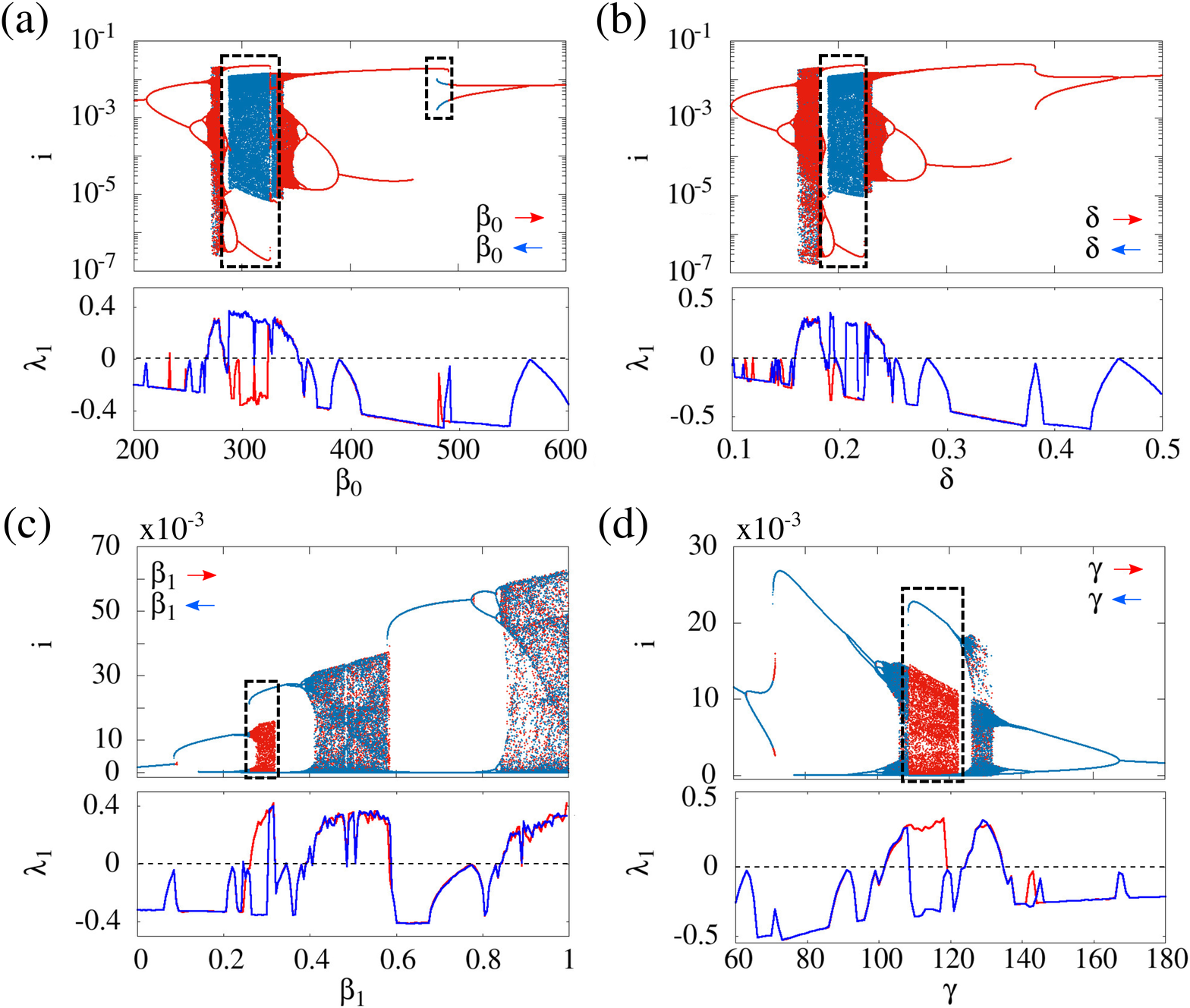}
\caption{Bifurcations diagram and respective largest Lyapunov exponent 
($\lambda_1$) for $\beta_0$ in the panel (a), for $\delta$ in
the panel (b), for $\beta_1$ in the panel (c), and for $\gamma$ in the panel
(d). The $i$ variable in the panels (a) and (b) is in log scale. We consider 
$\alpha=35.84$, $\mu=0.02$, $\omega=2\pi$, (a) $\gamma=100$, $\delta=0.20$,
$\beta_1=0.28$, (b) $\gamma=100$, $\beta_0=300$, $\beta_1=0.28$, (c)
$\gamma=100$, $\delta=0.25$, $\beta_0=270$, and (d) $\delta=0.25$,
$\beta_0=300$, $\beta_1=0.28$.}
\label{fig3}
\end{figure}

The influences of $\beta_0$, $\delta$, $\beta_1$, and $\gamma$
in the dynamical system are exhibited in Figs. \ref{fig3}(a-d),
respectively, by the bifurcation diagrams followed by the largest Lyapunov exponent ($\lambda_1$) \cite{wolf}.
The Lyapunov exponent is a tool to identify chaos \cite{tamastel}. A positive largest Lyapunov exponent
($\lambda_1 >0 $) corresponds to chaotic dynamics \cite{alligood}.
The bifurcations are constructed by the
collection of the $i$ maxima points in the forward (red) and backward (blue)
directions. The combination of distinct bifurcation in forward and backward
directions comprises a hysteresis \cite{everton}. Also, the Lyapunov exponents
are calculated in both directions. By considering these results, it is possible to locate
ranges where two attractors coexist, both looking at the bifurcation and 
Lyapunov exponent. These regions are delimited by the black dotted square 
in the panels (a-d). The regions where this dynamical behaviour 
exists are called bistability. Systems with these dynamics 
are extremely sensitive to the initial state \cite{feudel2008} and the 
presence of chaotic attractor in these systems is rare \cite{feudel1997, feudel2003}.

The result in Fig. \ref{fig3}(a) shows bistability dynamics in the
range $\beta_0\in[287,332]$ and $\beta_0\in[480,491]$. The bistability
comprehends 14\% of the $\beta_0$ range. However, for this parameter set, the
dynamics is mostly periodic. The bifurcation for $\delta$ is very similar to the
bifurcation for $\beta_0$, as shown in Fig. \ref{fig3}(b). For this
parameter set, the bistability between periodic and chaotic attractor only
exists in $\delta\in [0.19,0.22]$, which comprehends 7.5\% of the $\delta$ range.
Therefore, the time to lose immunity is relevant for the epidemic dynamics.
Diseases with long time immunity, for example $>10$ years, or short time 
immunity, as for example $<3$ years, have periodic dynamics. Another analised
parameter is the seasonality strength $\beta_1$, as displayed in Fig.
\ref{fig3}(c). Our simulations show one region of bistability in 
$\beta_1\in[0.25,0.32]$ (7\% of $\beta_1$ range) and most of the dynamics
is sustained by chaotic attractors. Figure \ref{fig3}(d) exhibits the
bifurcation as a function of $\gamma$, where one region of bistability in
$\gamma\in[108,123]$ (19\% of the $\gamma$ range), however, it is dominated by 
periodic behaviour. The $\omega$ parameter also is associated with the creation
or annihilation of bistability dynamics, for example, for $\omega$ in the
range 12 up to 30 months, the bistability is found. For values under 12
months the dynamics is periodic.

We observe that the parameters $\beta_0$, $\delta$, $\beta_1$, and $\gamma$ are
relevant to understand the disease spread dynamics. However, we select the
$\alpha$ parameter as the control parameter, once the bifurcation
diagram for this parameter exhibits  rich dynamic, it is possible to study the
crisis and critical points, known as tipping points. 


\section{Bifurcation analysis}

The latent period is an important variable in the dynamics of the epidemics
\cite{lessler2009} and is defined as $1/\alpha$ \cite{rbef}. To understand the
effects of $\alpha$ in the dynamical system (Eq. \ref{eq3}), we consider
$\delta=0.25$, $\beta_0=270$, $\beta_1=0.28$, $\gamma=100$, and $\mu=b=0.02$,
where the time unity is year. We choose $\alpha$ as the control due to the fact that
the bistability and tipping points are more evident.

\begin{figure}[hbt]
\centering
\includegraphics[scale=0.25]{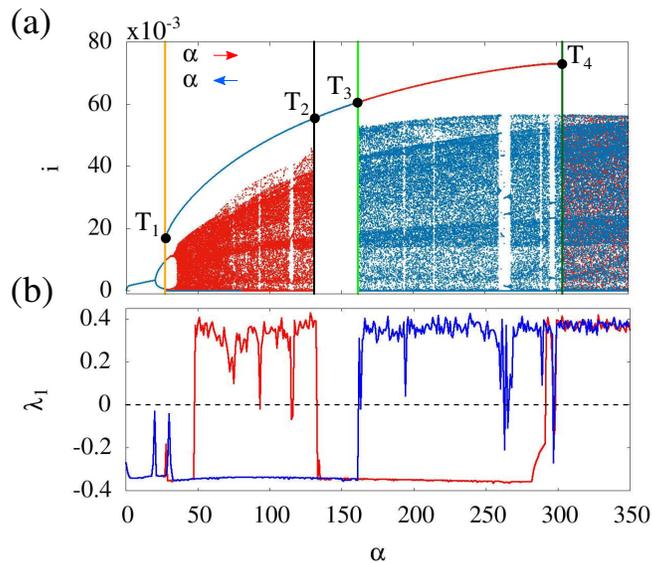}
\caption{(a) Bifurcation diagram and (b) largest
Lyapunov exponent. The red points and lines are forward and the
blue points and lines are backward in the $\alpha$ parameter. We consider $T_1=27$,
$T_2=130.88966$, $T_3=161.69990$, $T_4=303.88407$, $\delta=0.25$, $\beta_0=270$,
$\beta_1=0.28$, and $\gamma=100$.}
\label{fig4}
\end{figure}

Figure \ref{fig4} displays the bifurcation diagram (considering
$i$ maximum) in the panel (a) and the largest Lyapunov exponent ($\lambda_1$)
in the panel (b).
Given an initial condition, the red point is the maximum value of $i$ in
the forward $\alpha$ direction following the attractor, i.e., the initial
condition for the current step is equal to the last step. The blue 
points are
obtained in the backward $\alpha$ direction. In Fig. \ref{fig4}(a), 
the vertical lines ($T_i$) are the
critical points that delimit the bifurcation, where $i = 1,2,3,4$.  The
bistable chaotic-periodic occupies $\approx 70 \%$ of the $\alpha$ range, while
the coexistence between periodic-periodic $\approx 1\%$. The 
ranges $[T_1,T_2]$ and $[T_3,T_4]$ encompass the coexistence 
of $\lambda_1<0$ and $\lambda_1>0$ which confirms the bistability between
periodic and chaotic attractors, as illustrated in Fig. \ref{fig4}(b).

\begin{figure}[hbt]
\centering
\includegraphics[scale=0.14]{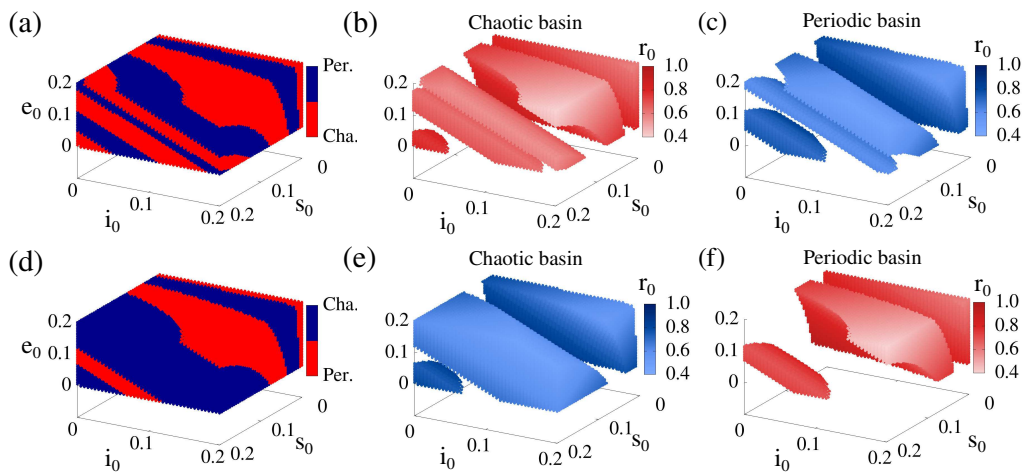}
\caption{Basin attraction calculated for $\alpha=100$. In panel (a) there is
the total basin attraction, in panel (b) the basin of chaotic attractor, and in
panel (c) the periodic basin. In panels (d), (e), and (f), we calculate
the basin for $\alpha=200$. We compute the chaotic and periodic basins in the
panels (e) and (f), respectively.}
\label{fig5}
\end{figure}

Due to the coexistence of two attractors, we compute two basins of
attractions for different $\alpha$ values. For $\alpha = 100$ in 
Figs. \ref{fig5}(a-c) and for $\alpha=200$ in Figs. \ref{fig5}(d-f).
The colour scheme follows the colour of the attractor in Fig.~\ref{fig4}. 
For $\alpha=100$, 43\% of the basin formed by red points
that evolve to the chaotic attractor is separately displayed in Fig.~\ref{fig5}(b),
while 57\% of the basin is composed of blue points, which  evolve to the periodic attractor
as displayed in  Fig.~\ref{fig5}(c). The colour scale in Figs.~\ref{fig5}(b), 
 \ref{fig5}(c), \ref{fig5}(e) and \ref{fig5}(f) is $r_0$. The composition of the basin attraction is not
preserved by $\alpha$ translation. However, in other
ranges, for example $[T_3,T_4]$, the shape of the basin attraction  changes, as shown
in Fig.~\ref{fig5}(d), for $\alpha = 200$. For this
$\alpha$ value, 57\% of the points evolve to the chaotic attractor.
The basin for the chaotic attractor is shown in Fig.~\ref{fig5}(e). 
The 43\% of the remaining points evolve to the periodic attractor and are plotted
in Fig.~\ref{fig5}(f). The structure of this basin remains in the range
$[T_3,T_4]$. However, translations in $\alpha$ change the composition 
of the basin attraction, which decrease as a cubic function.

The sudden change in the dynamical behaviour occurs at certain values of the
control parameter, that are the critical points $T_1=27$, $T_2=130.88966$,
$T_3=161.69990$, and $T_4=303.88407$. These events are called crises
\cite{grebogi1982, grebogi1983} and are defined by the collision between a
chaotic and a periodic attractor. 

In the point $T_1$ occurs a rising of a bistable region by bifurcation. It
starts by the existence of two different attractors, that are represented by 
blue and red branches. The bistability is formed by two periodic
attractors until the point in which the red branch becomes chaotic by
bifurcation. After that, the dynamics is sustained by the coexistence of periodic
(blue) and chaotic (red) attractors until $T_2$.

\begin{figure*}[hbt]
\centering
\includegraphics[scale=0.37]{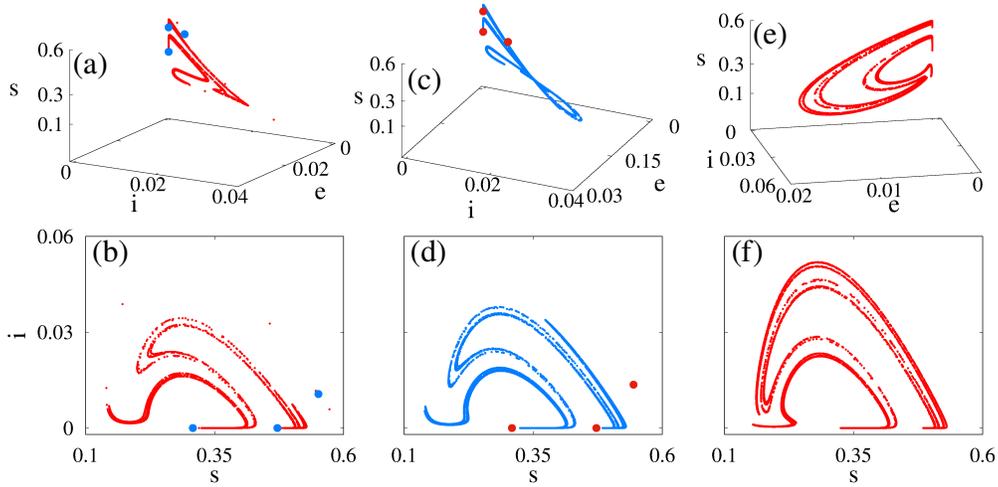}
\caption{Coexistence of periodic and chaotic attractors. Three dimensional
projection in the panels (a), (c), and (e). Bidimensional projection in the
panels (b), (d), and (f). We consider $\alpha=130$ ((a) and (b)),
$\alpha=161.70276$ ((c) and (d)), and $\alpha=303.88411$ ((e) and (f)). In the
panels (a) and (b), the initial conditions for periodic is  
$(s_0,e_0,i_0,r_0)=(0.074,0.144,0.186,0.596)$ and for chaotic is
$(s_0,e_0,i_0,r_0)=(0.075,0.145,0.185,0.595)$. For the panels (c) and (d), the
initial conditions for periodic and chaotic are
$(s_0,e_0,i_0,r_0)=(0.074,0.144,0.186,0.595)$ and
$(s_0,e_0,i_0,r_0)=(0.060,0.100,0.110,0.730)$, respectively. For the panels (e)
and (f) are $(s_0,e_0,i_0,r_0)=(0.074,0.144,0.186,0.596)$ and 
$(s_0,e_0,i_0,r_0)=(0.031,0,026,0,073,0,870)$. We consider $\mu=0.02$,
$\omega=2\pi$, $\gamma=100$, $\delta=0.25$, $\beta_1=0.28$, and $\beta_0=270$.}
\label{fig6}
\end{figure*}

The $3$-dimensional and $2$-dimensional projections of the attractors merging in
$[T_1,T_2]$ range are displayed in Figs. \ref{fig6}(a) and \ref{fig6}(b),
respectively, for $\alpha = 130$. The attractor shape is invariant by
$\alpha\in [T_1, T_2]$ translation. The attractors are constructed considering
the stroboscopic map, that is a collection of the dynamical variables every
$T=n\pi$, where $n=0,2,4,...$. Figures \ref{fig6}(c) and \ref{fig6}(d) display
the $3$ and $2$-dimensional projections of the chaotic (blue points) and
periodic (red points) in the range $[T_3,T_4]$ for $\alpha=161.70276$. Figures
\ref{fig6}(e) and \ref{fig6}(f) show the projections of the chaotic attractor
for $\alpha>T_4$. In this regime, only chaotic attractor survives.

\begin{figure}[hbt]
\centering
\includegraphics[scale=0.5]{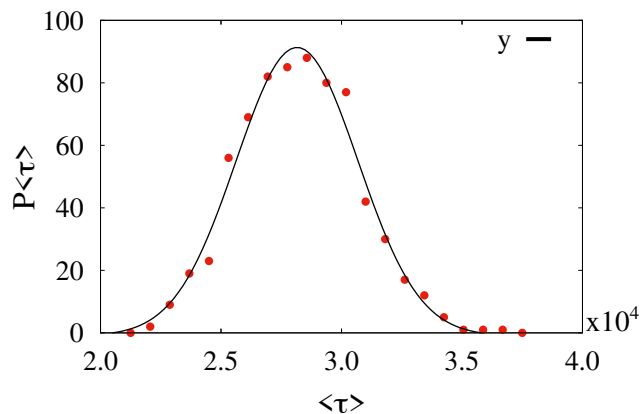}
\caption{Distribution for the average transient time $\langle\tau\rangle$ in red
points and range $[T_2,130.89666]$. The average is calculated for $200$ samples
of chaotic initial conditions. The black curve is a gaussian fit.
We consider $\mu=0.02$, $\omega=2\pi$, $\gamma=100$, $\delta=0.25$,
$\beta_1=0.28$, and $\beta_0=270$.}
\label{fig7}
\end{figure}

Once crossed $T_2$, the bistability disappears and periodic behaviour prevails.
In this point, the chaotic attractor collides with the periodic one by a 
saddle-point bifurcation.  Around the crisis point, it is expected that
the transient time tends to infinity. We calculate the average transient time
$\langle\tau\rangle$ for $200$ initial conditions in the interval
$[T_2,130.89666]$, as displayed in Fig. \ref{fig7} by red points. $\tau$ 
parameter represents the time to the chaotic attractor dynamics goes to 
the periodic one. The black
curve is a Gaussian fit displayed by the equation 
\begin{equation}
y=y_{0}+\frac{A}{\sigma\sqrt{\frac{\pi}{4\ln(2)}}}e^{-\frac{4\ln(2)(\langle\tau\rangle-\langle\tau\rangle_{\rm c})^{2}}{\sigma^{2}}}.
\end{equation}
The $y_{0}$, $A$, $\sigma$ are parameters for the fitting. We observe the existence of a transient chaos around $T_2$, that has a
distribution type Gaussian centered in $\langle\tau\rangle_{\rm c}=28\times 
10^4$ years. 
In the range where transient chaos exists, the basin of attraction is composed of chaotic and
periodic orbits. However, the initial conditions that evolve to chaotic
attractor dissipate linearly and smoothly with the increase of the transient
time.

\begin{figure}[hbt]
\centering
\includegraphics[scale=0.5]{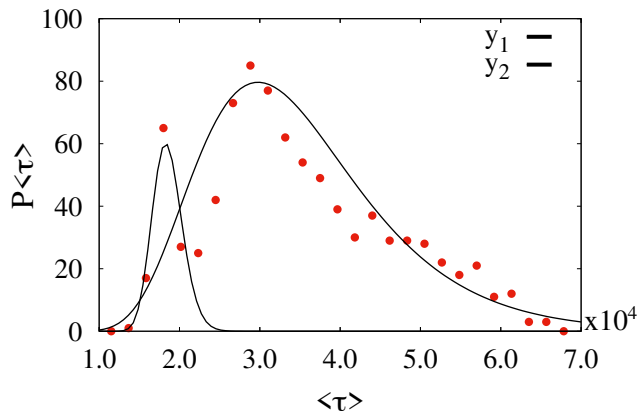}
\caption{Distribution for the average transient time $\langle\tau\rangle$ in
$[T_3,161.70849]$ (red points). The black curve is the fit given by a log normal. The average is
calculated for $200$ samples. We consider $\mu=0.02$, $\omega=2\pi$,
$\gamma=100$, $\delta=0.25$, $\beta_1=0.28$, and $\beta_0=270$.}
\label{fig8}
\end{figure}

The point $T_3$ denotes the birth of a new bistable region. That happens by a
transition from a periodic branch (blue) to a chaotic by a saddle point. In
this transition also exists a transient chaos, which is 
displayed in Fig. \ref{fig8} by the red points. The transient time in the range
$[T_3,161.70849]$ follows a binomial distribution, indicate by the 
continuous black lines $y_1$ and $y_2$, given by the log normal 
distribution
\begin{equation}
y=y_0+\frac{A}{\sqrt{2\pi} \sigma \tau}e^{-\frac{({\ln}( 
\langle\tau\rangle / \langle\tau\rangle_{\rm c}))^2}{2\sigma^2}},
\end{equation}
where the first peak value is $\langle\tau\rangle=1.9\times 10^4$ years and the
second one is $\langle\tau\rangle=3\times 10^4$ years. For $\alpha<T_3$ and
close to $T_3$, there is transient chaos. The basin of attraction for the chaotic
attractor is extinct linearly smooth with the increase of the transient. 

\begin{figure}[hbt]
\centering
\includegraphics[width=0.6\textwidth]{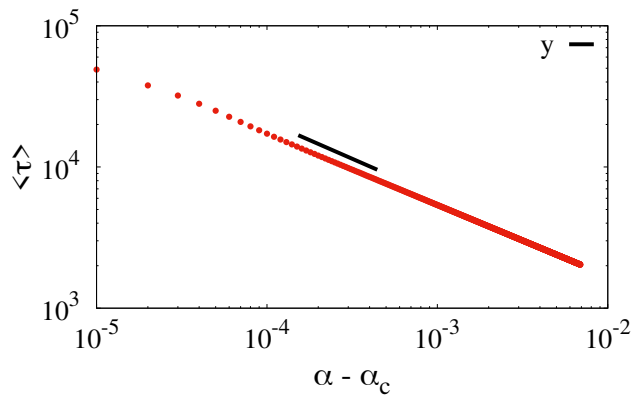}
\caption{Average transient lifetime versus $\alpha-\alpha_{\rm c}$ in 
log scale for $200$ periodic initial conditions (red points). The black line
indicates the fitting. We consider $\mu=0.02$,
$\omega=2\pi$, $\gamma=100$, $\delta=0.25$, $\beta_1=0.28$, and $\beta_0=270$.}
\label{fig9}
\end{figure}

After crossing $T_3$, we observe a bistable regime that comprehends 40\% of the
$\alpha$ range. The bistability is sustained by the coexistence of periodic and 
chaotic attractors, as shown in Figs. \ref{fig6}(c) and \ref{fig6}(d). Crossing
this bistable regime, we find the last crisis point ($T_4$). The transient 
time goes to infinity, following a decay $\propto x^{-1/2}$ as we move away from
$T_4$. However, at this point, the transient is periodic. Figure \ref{fig9}
exhibits our result for $200$ periodic initial conditions. Differently from the
two first cases, the transient goes to infinity for $T_4=\alpha_c$ and decays
with $(\alpha-\alpha_c)^{\gamma}$, where $\gamma 
= -0.5025 \pm 0.0002$ and $\alpha>\alpha_c$, with this standard 
deviation value we can say that $\gamma = 1/2$ that is the universality 
exponent for average duration of chaotic transients. The black line 
indicates the fit curve, that is given by  $y=167.06(\alpha-\alpha_c)^{-0.5025}$ with a
correlation coefficient equal to $-0.9999549$. After this transient,
the periodic points, in the phase space, coalesce in a chaotic attractor by a
saddle-point. The basin of attraction in the crisis point is formed by 19\% of the
initial conditions that evolve to a periodic attractor. Increasing the transient
time until $\tau=50800$, we verify an abrupt change and the fraction of periodic
points goes to zero, discontinuously. The periodic transient persists until
1\% above $\alpha_c$ calculated through $(\alpha-\alpha_c)/\alpha_c$.

\section{Tipping points}

Tipping points mark changes in the system between alternative states
\cite{dakos2019}. It occurs when a threshold is crossed due to an external
perturbation or by shift in the parameters of the system \cite{vannes2016}.
The transitions correspond to saddle-node or fold bifurcations \cite{everton}.
Between two tipping points, the system is bistable, i.e., it can be found in
one of two possible states \cite{dakos2019}. 

In the seasonal disease context, the desired state is one in which the future
outbreak is predictable. If the disease spread is predictable, based on the
data from previous years, it is possible to construct more efficient control
strategies and, consequently, decrease the number of infected individuals. The
undesired state, on the other hand, is when the evolution of the disease spread
is unpredictable. The unpredictability is associated with chaotic 
dynamics
\cite{stollenwerk2012} and was studied by Scarpino and Petri
\cite{scarpino2019}. However, until the moment, the mechanism behind the
unpredictability is not satisfactorily explained by only the chaotic dynamics.
In this work, we show that unpredictability is associated with 
bistable dynamics and has a tipping point.

\begin{figure}[hbt]
\centering
\includegraphics[width=0.7\textwidth]{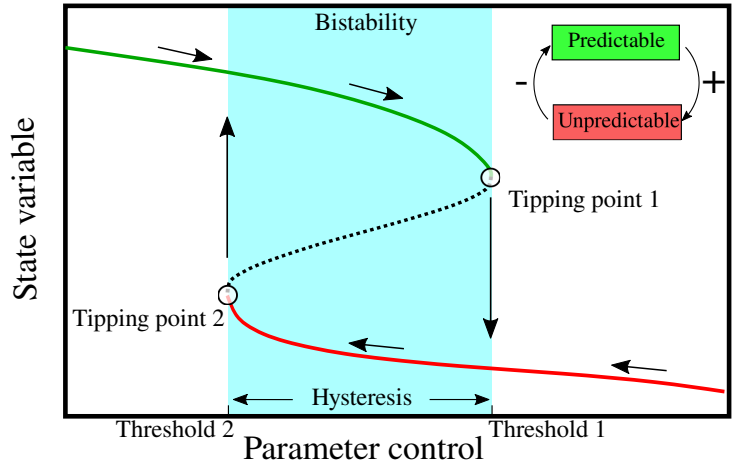}
\caption{Representation of the tipping points between predictable 
(green branch)
and unpredictable (red branch) states.}
\label{fig10}
\end{figure}

Figure \ref{fig10} exhibits the state variable as a function of the parameter
control. We observe a state related to the predictable (green line) and another
to the unpredictable (red line). The state variable can be the number of
infected and the parameter control can be the $\alpha$ variable. With regard to
the green curve, the control parameter is increased  until tipping point 
$1$. At this point, the system reaches threshold $1$. Once crossed, the
state variable evolves to the red branch, which represents the unpredictable
state. If we start in the red branch and decrease the control parameter, then,
after crossing threshold $2$, the system shifts to the green branch.
Therefore, the state system can alternate between unpredictable and predictable due to parameter control.
This situation illustrates what happens in points $T_3$ and $T_4$, as shown
in Fig. \ref{fig4}.

Firstly, we focus on the range $[T_1,T_2]$ shown in Fig. \ref{fig4}. The system
exhibits periodic behaviour for $\alpha<T_1$ and $\alpha>T_2$. However,
it encompasses a bistable chaotic-periodic dynamics in the considered 
interval. In this region, the maximum number of
infected individuals increases by 20 times from $T_1$ to $T_2$, 
however, in the boundary the attractors are periodic. The 
bistable region is interesting in terms of predictability, for the reason that 
small changes in the initial condition can leave the system from periodic to
chaotic behaviours. In this region, the average probability of an initial
condition evolving to the periodic behaviour is 49\%. Therefore, with this
measure, the predictability is uncertain in the
interval $[T_1,T_2]$.

\begin{figure}[hbt]
\centering
\includegraphics[width=0.65\textwidth]{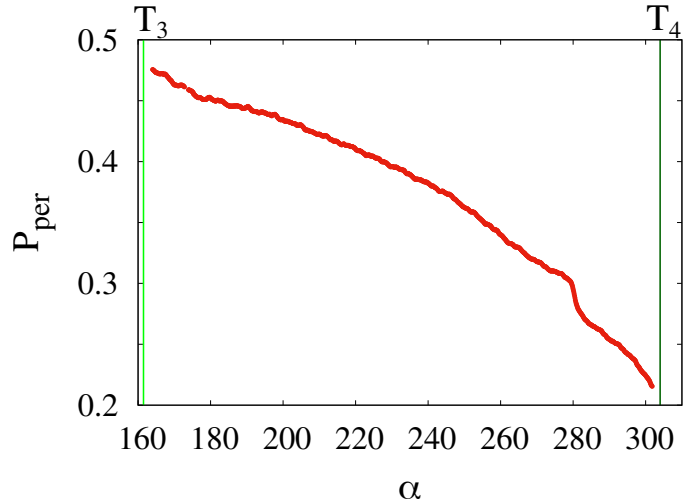}
\caption{The probability of an initial condition evolves to the periodic attractor in
the range $[T_3,T_4]$ (vertical lines). We consider $\mu=0.02$, $\omega=2\pi$,
$\gamma=100$, $\delta=0.25$, $\beta_1=0.28$, $\beta_0=270$.}
\label{fig11}
\end{figure}

In the range $[T_3,T_4]$ are the tipping points illustrated in Fig. \ref{fig10}.
Considering $T_2<\alpha<T_3$, all the initial conditions evolve to a periodic
attractor, namely the dynamics is predictable. For example, based on the date of
one year, it is possible to implement restrictions for the next year. Once
crossed $T_3$, the dynamic becomes bistable. The range $[T_3,T_4]$ is connected
by the coexistence of chaotic and periodic dynamics. The probability of one
initial condition evolving to a periodic attractor is displayed in Fig.
\ref{fig11}. The probability distribution for $\alpha$ has a cubic decay as
closer to $T_4$. The average is equal to 37\% and we can affirm that diseases in
$[T_3,T_4]$ are preferably unpredictable. After crossing $T_4$, all the initial
conditions evolve to the chaotic attractor and, as a consequence, the disease spread becomes unpredictable. 

\section{Conclusions}

In this work, we study a SEIR seasonal model with temporary immunity. The
inclusion of temporary immunity greatly enriches the system dynamics. Our
results show that the bistable dynamics depend on the control 
parameters. Thus, by varying $\alpha$, 70\% of the range exhibits
bistability, which is composed of chaotic and periodic attractors. Our results
show the importance of all parameters in the spread dynamics, however, with the
crisis present in the $\alpha$ bifurcation, it is possible to study tipping
points phenomena. We explore a range that includes diseases with a latent period
in order of days until hours \cite{lessler2009}.

We verify that the dynamics of the disease spread is chaotic for
$\alpha \geq 300$. The diseases in this range have a latent period less than
$1.2$ days. Only values in $0 < \alpha \leq 30$ (very high values of
$1/\alpha$ until $12$ days) and $130 \leq \alpha \leq 161$ ($2.8$ until $2.2$
days) are periodic. Values above $\alpha>350$ are not considered.

The analysed range shows that the latent period is a crucial variable to
understand the reason for the unpredictability of infectious diseases. This
unpredictability was observed by Scarpino and Petri  \cite{scarpino2019},
however, the components of the unpredictability were unclear. Our results show
that the unpredictability is closely associated with $\alpha$ due to bistable
dynamics.

To finish this work, we answer the question provided in Introduction: the
disease spread becomes unpredictable when the tipping point $T_4$ is crossed.
Therefore, diseases with a short latent period, less than 30 hours, are always
unpredictable. 


\section*{Acknnowledgements}

The authors thank Dr. E.S. Medeiros for discussions.
 The authors thank the financial support from the Brazilian Federal 
 Agencies (CNPq), grants 407299/2018-1, 311168/2020-5, the S\~ao Paulo Research
Foundation (FAPESP, Brazil) under grants 2018/03211-6, 2020/04624-2, 2022/05153-9, 
CAPES, Funda\-\c c\~ao A\-rauc\'aria.
The authors thank 3D NeuroNets LLC. We thank 105 Group Science (www.105groupscience.com).


\end{document}